\documentclass[10pt,letterpaper]{article}

\usepackage{cogsci}

\cogscifinalcopy % Uncomment this line for the final submission 

\usepackage{pslatex}
\usepackage{apacite}
\usepackage{float} % Roger Levy added this and changed figure/table
\usepackage{natbib}
\usepackage{graphicx}
\usepackage{multirow}
\usepackage{amsmath} 
\usepackage{subcaption}
\usepackage{tikz}

\title{MPPFND: A Dataset and Analysis of Detecting Fake News with Multi-Platform Propagation}

\author{
{\large \bf Congyuan Zhao$^{1,2}$, Lingwei Wei$^{1*}$, Ziming Qin$^{1,2}$, Wei Zhou$^{1}$, Yunya Song$^{3}$, Songlin Hu$^{1,2*}$} \\
$^1$Institute of Information Engineering, Chinese Academy of Sciences \\
$^2$School of Cyber Security, University of Chinese Academy of Sciences \\
$^3$Division of Emerging Interdisciplinary Areas, Hong Kong University of Science and Technology \\
\{zhaocongyuan, weilingwei, zhouwei, husonglin\}@iie.ac.cn, qinziming21@mails.ucas.ac.cn, yunyasong@ust.hk
}

\begin{document}
\maketitle
\begin{abstract}

Fake news spreads widely on social media, leading to numerous negative effects. Most existing detection algorithms focus on analyzing news content and social context to detect fake news. However, these approaches typically detect fake news based on specific platforms, ignoring differences in propagation characteristics across platforms. In this paper, we introduce the MPPFND dataset, which captures propagation structures across multiple platforms. We also describe the commenting and propagation characteristics of different platforms to show that their social contexts have distinct features. We propose a multi-platform fake news detection model (APSL) that uses graph neural networks to extract social context features from various platforms. Experiments show that accounting for cross-platform propagation differences improves fake news detection performance.

% \iffalse
% The proliferation of fake news across various social media platforms has led to numerous negative consequences. Most existing fake news detection algorithms focus on analyzing both news content and social context to identify fake news. However, current methods for collecting social context often standardize the data across different platforms, overlooking the differences in the propagation characteristics of these platforms. In this paper, we first introduce a benchmark MPPFND fake news dataset annotate with platform-specific labels. Additionally, we provide a comprehensive description of the commenting and propagation characteristics across different platforms, confirming that social context varies significantly among platforms. We further propose an effective multi-platform fake news detection model (APSL) that employs various graph neural networks to extract social context features from different platforms. Experiments demonstrate that the multi-platform fake news detection model (APSL) can improve the performance of fake news detection.
% \fi

\end{abstract}

\section{Introduction}

% \iffalse
% 虚假信息检测简要背景
% (虚假新闻定义)
% \fi

Fake news is defined as fake news that is deliberately created and disseminated, typically with the intent to mislead\citep{fake_news_defination}. The widespread propagation of fake news on social platforms has had significant negative impacts on society. For instance, during the COVID-19 pandemic, a plethora of fake news caused widespread panic among the public, and incorrect treatment methods adversely affected many lives \citep{intro_1.2,intro_1.3}. Many fact-checking sites\footnote{Such as politifact.com, snopes.com etc.} are dedicated to having fact-checkers collect accurate information and compare it with news appearing on the internet to identify fake news. However, manual verification is too inefficient to keep up with the vast amount of fake news. Therefore, there is an urgent need for an effective model to automatically detect fake news and curb the spread of fake news.
\renewcommand{\thefootnote}{\fnsymbol{footnote}} % 设置脚注符号为 *, †, ‡ 等
\footnotetext[1]{Corresponding author.}
\renewcommand{\thefootnote}{\arabic{footnote}}   % 恢复正常编号（1, 2, 3...）

% \iffalse
% 简要介绍一下虚假信息检测的发展
% 我的思路是从特征选取的角度出发：
% 基于文本信息的研究，如众包专家知识库的检测方法和FakeBert一类的模型
% 另一部分研究加入了传播结构特征，如GCNFN和GNN-CL
% 后来部分研究关注文本特征的多源性，出现了多域模型，以MCFEND一众文章为主
% 目前对于传播结构的多源性的研究还有所缺失，本文认为不同平台的传播结构特性不同且该特性有助于对虚假和真实信息进行分类。

% 目前虚假新闻检测方法从特征选取的的角度出发主要有两类：文本特征和传播结构特征。基于文本特征的模型如FakeBert，旨在通过新闻源文本之间的语言差异性来检测虚假信息。分析包括写作风格，情感偏好等。近期以MCFEND系列研究为代表开始对新闻文本进行更细粒度的分析，划分出了不同的事件域来进行多域的虚假新闻检测。基于传播结构的模型如GAMC的模型则将新闻根据其在社交媒体上的传播过程构建为传播树从而将问题转化为图分类问题。之后如RAGCL的工作进一步讨论了传播树的结构特征。但目前的研究在细粒度的传播结构研究上有所缺失，我们认为不同平台的社交媒体信息上的传播过程会携带不同的传播特征，因此我们认为对传播结构进行更精细的分析和建模是有必要的。
% (基于文本 基于传播 
% 分段 但是都是基于单平台，尽管有一些工作如MCFEND)
% （基于内容 融合传播结构：基于序列、基于图）

% \fi

% Currently, many studies mainly use content features and propagation structure to detect fake news: Content feature based models \citep{writing_style_2_FakeBERT, eann, ma2017content, jin2017detection} aim to detect fake news through language differences between news contents. Recent studies have explored more fine-grained analyses of news content, such as using multi-domain models\citep{domain_2_MDFEND, mcfend} that categorize news into different event domains for multi-domain fake news detection. Some works \citep{propagation_GNN_GAMC, bigcn, propagation_GNN_RAGCL} incorporate user interactions such as retweets and comments during the propagation. These research model them as a propagation sequence\citep{seq_model1, seq_model2} or graph\citep{graph_model1, graph_model2} to learn sequential or topological structures for better detection.

Currently, many studies mainly use content features and propagation structure to detect fake news: Content feature based models \citep{writing_style_2_FakeBERT, eann, ma2017content, jin2017detection} aim to detect fake news through language differences between news contents. Recent studies have explored more fine-grained analyses of news content, such as \citep{domain_2_MDFEND, mcfend} that categorize news into different event domains for  fake news detection. Some works incorporate user interactions such as retweets and comments during the propagation. These research model the spread of news as propagation sequence \citep{seq_model1, propagation_attention_PLAN} or graph \citep{domain_1, propagation_GNN_GAMC, propagation_GNN_GAMC, bigcn, propagation_GNN_RAGCL} to learn sequential or topological structures for better detection.

% \iffalse

% In the real word, a piece of fake news often spreads across multiple platforms through mechanisms such as retweets. Multiple platforms often provide richer and complementary features for news detection. However, most works usually detect fake news based on the specific platform and obtain limited performance due to partial modeling. We believe that the content process on social media information on different platforms will carry different propagation characteristics. Therefore, we believe that a more detailed analysis and modeling of the propagation structure is necessary.

% 在Hilde A. M的研究中，探讨了平台类型差异化的作用。该文章详细论述了不同社交平台间互动信息的差异性。
% （增加具体新闻示例）
% 因此我们认为对不同平台的传播差异进行自适应的建模有助于提升虚假信息检测系统的性能。因此我们收集了包含不同社交媒体平台传播结构信息的数据集并提出了自适应建模不同平台信息传播差异的模型来验证我们的猜测。

% In the study of  \citep{intro_1.4}, the role of platform type differentiation was explored. The article discusses in detail the differences in interactive information between different social platforms. We believe that adaptive modeling of the propagation differences between different platforms will help improve the performance of fake news detection systems. Therefore, we collected a dataset containing information about the propagation structure of different social media platforms and proposed a model for adaptively modeling the differences in information propagation between different platforms to verify our hypothesis.

% \fi

\begin{figure}[t]
  \centering
  \includegraphics[width=0.9\columnwidth]{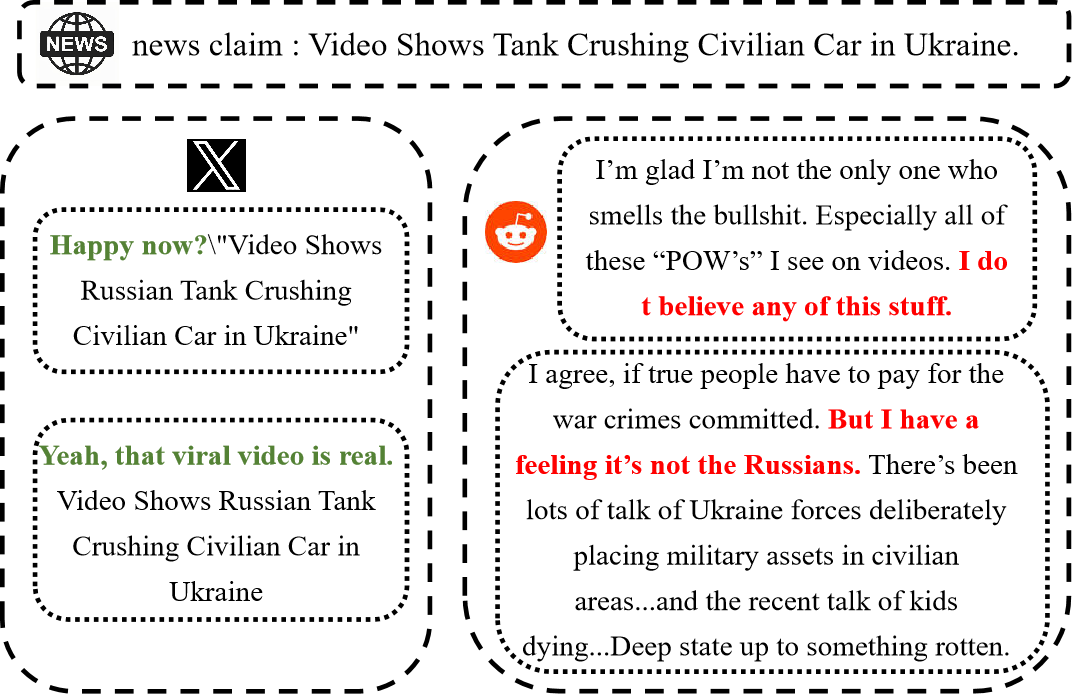}
  \caption{An example of real-world news spreading across multiple social media platforms, where we can observe a significant disparity in how the credibility of the news is assessed by the two platforms.}
  \label{fig:multi-propagation-news-exp}
\end{figure}

\begin{table*}
  \centering
  \resizebox{0.98\textwidth}{!}{
    \begin{tabular}{ccccll}
    \hline 
        \multicolumn{1}{c}{\multirow{2}[3]{*}{Dataset}} 
        & \multicolumn{3}{c}{\# Feature} 
        & \multicolumn{1}{c}{\multirow{2}[3]{*}{\# Social Platform Source}} 
        & \multicolumn{1}{c}{\multirow{2}[3]{*}{\# Language}} 
        \\
        \cline{2-4}          
        & \multicolumn{1}{c}{Content} & \multicolumn{1}{c}{Single-platform} & \multicolumn{1}{c}{Multi-platform} &         \\
        &  & Propagation & Propagation 
        \\
        \hline
        \multicolumn{1}{l}{FakeNewsNet \citep{fakenewsnet}} & \multicolumn{1}{c}{
        $\surd$} &  \multicolumn{1}{c}{
        $\surd$} & ${\times}$  & {X}  & {EN} \\
        \multicolumn{1}{l}{Weibo16 \citep{weibo16}} & \multicolumn{1}{c}{
        $\surd$}  & \multicolumn{1}{c}{
        $\surd$} & ${\times}$ & {Weibo}  & {CH} \\
        \multicolumn{1}{l}{Weibo21 \citep{domain_2_MDFEND}} & \multicolumn{1}{c}{
        $\surd$} & \multicolumn{1}{c}{
        $\surd$} & ${\times}$  &  {Weibo}  & {CH}  \\
        \multicolumn{1}{l}{Twitter16 \citep{twitter16}}  & \multicolumn{1}{c}{
        $\surd$} & \multicolumn{1}{c}{
        $\surd$} & ${\times}$  &  {X}  & {EN}  \\
        \multicolumn{1}{l}{MCFEND \citep{mcfend}} & \multicolumn{1}{c}{
        $\surd$} & \multicolumn{1}{c}{
        $\surd$} & ${\times}$ & {Weibo} & {CH}  \\
        % \hline
        \multicolumn{1}{l}{\textbf{MPPFND} (Ours)}  & \multicolumn{1}{c}{
        $\surd$} & \multicolumn{1}{c}{
        $\surd$} & $\surd$ & {X\textbackslash{}Youtube\textbackslash{}Reddit} & {EN}  \\
        % &       &       &       &       &  \\
        \hline 
        \end{tabular}%
    }
  \caption{
    Comparison of our dataset with other popular fake news detection datasets.
  }
  \label{tab:dataset-feature}
\end{table*}

However, most works usually detect fake news based on the specific platform and obtain limited performance. 
Some works \cite{intro_1.4}  have shown the difference between various social media platforms.
% In the study by  \citep{intro_1.4}, the article highlighting the difference between various social media platforms.
In the real world, a piece of fake news often spreads across multiple platforms through mechanisms such as retweets, that often provide richer and complementary features for detection~\ref{fig:multi-propagation-news-exp}. Therefore, a more detailed analysis and modeling of the propagation structure is necessary.

In this paper, to fill the gap in this area, we aim to conduct an in-depth analysis of the characteristics of multi-platform propagation and explore that these differentiated features aid in detecting fake news. We collect a new fake news dataset that gathers the propagation structure of a news claim across various social media platforms. We list existing popular fake news datasets below and compare them with our dataset repository in Table~\ref{tab:dataset-feature}. Based on our constructed dataset, we further analyze the propagation characteristics across platforms, revealing significant variations in both commonalities and differences for fake news detection. 

By analyzing our dataset, we find that user engagement effectively differentiates between the authenticity of news on YouTube and X, whereas this feature does not show a significant relationship with news authenticity on Reddit. Additionally, the method of using comment sentiment to distinguish news authenticity is applicable to X(Twitter) and Reddit, but shows only moderate effectiveness on YouTube. Thus, it is evident that differential modeling across various platforms is necessary for detecting fake news.

To demonstrate that the platform-specific characteristics we identified contribute to the detection of fake news, we propose a multi-platform detection model that leverages multiple graph neural networks to extract features from propagation structures across different platforms. Our experiments show that the differences in propagation across platforms contribute to improving the effectiveness of fake news detection models.

% \iffalse
% 本篇论文的贡献在于：
% a.我们提出了MPPFND数据集（multi-platform propagation dataset）：一个包含不同社交媒体平台传播结构信息的英文虚假信息检测数据集。该数据集包含4965条claim，来自两个事实核查网站，同时还收集这些新闻在三个不同的社交媒体平台上的传播情况。我们在该数据集上进行了统计，分析了不同社交媒体平台的传播内容差异。（数据集条目，来源 信息 贡献）
% b.我们提出了一个自适应建模不同平台信息传播差异的模型：Adaptive Multi platform fake news detection Model，并在我们提出的数据集和mcfend数据集上进行了测试，测试的结果表明包含不同平台的传播结构信息有助于对虚假新闻进行检测。（分析发现传播存在差异）
% c.通过对实验结果的分析，我们验证了不同平台信息传播差异有助于对虚假新闻进行检测，这给虚假信息检测提供了新的思路。
% \fi

The contribution of this paper lies in:
% \begin{itemize}
%     \item 
    % We build the MPPFND dataset (Multi-Platform Propagation Dataset for Fake News Detection): an English fake news detection dataset that contains information on the propagation structure of different social media platforms. This dataset collects the propagation of these news on three different social media platforms.
   1) We construct and release a new MPPFND dataset for fake news detection with multi-platform propagation. It contains 3,500+ claims, along with 440,000+ user engagements across 3 mainstream social platforms.\footnote{We release our dataset and code at \\ https://github.com/Zhaocongyuan/MPPFND-Dataset.}
    % The dataset is derived from different hot events and contains  about 3,500 claims and the corresponding propagation across different types of social media platforms. 
% 
    % \item 
    % We conduct a thorough analysis of the propagation characteristics across different platforms and find that features such as user engagement and comment sentiment impact news authenticity differently across various social media platforms. Furthermore, we reveal the significance of these features in affecting the authenticity of news on specific platforms.
    2) We contribute empirical insights into propagation patterns across different social platforms for fake news detection.
    Our findings reveal platform-specific and platform-shared propagation patterns on news content, comment style, and user engagement for fake news detection.
    % Our findings reveal the propagation preferences of different platforms and the Echo Chamber Effect present on platforms. Additionally, we analyzed the features that can assist specific platforms in distinguishing the authenticity of news. 
    % \item 
    % We build an adaptive model for modeling the differences in information propagation across different platforms: the Adaptive Multi platform fake news detection Model(APSL), and test it on our dataset. The test results show that including different on the propagation structure of various platforms helps to detect fake news.
    3) We evaluate existing content- and propagation-based fake news detection methods for multi-platform fake news detection and further develop an adaptive propagation structure learning network (APSL) to capture platform-adaptive structural features from propagation across platforms for better detection.
    % and discuss some potential research directions.
% \end{itemize} 

\section{Related Work}

% 虚假信息检测
% 2.1-Content-Based Methods
% 由于新闻内容的丰富性，基于内容的检测方法旨在找到虚假新闻中所共有的模式与特征信息，如语言特征 \citep{r2.1.1, r2.1.2}、写作风格 \citep{r2.1.3, r2.1.4}和情感特征 \citep{r2.1.5, r2.1.6, r2.1.7}。

% 然而，基于内容的方法面临以下挑战。首先，新闻信息往往具有较强的时效性，导致模型学习到的知识不能完全泛化到新出现的新闻 \citep{r2.1.8}。其次，大型语言模型(LLMs)的出现大大降低了生成假新闻和改写人造假新闻的门槛，这导致通过内容特征来鉴别假新闻更加困难 \citep{r2.1.9}。
% 虚假信息数据集

\subsection{Fake News Detection}
Previous works on fake news detection can be broadly categorized into content-based and propagation-based methods.

\noindent\textbf{Content-Based Methods.} Content-based detection identifies recurring patterns and distinctive features in fake news by analyzing news articles \citep{acl_content_2}. This includes examining linguistic traits \citep{linguistic_1}, unique writing styles \citep{writing_style_1, writing_style_2_FakeBERT}, and emotional nuances \citep{emotional_1, emotional_2, emotional_3}. Recent studies have also focused on domain-specific features \citep{domain_1, domain_2_MDFEND}. Additionally, some approaches extend the analysis to multimodal elements, such as images \citep{acl_content_1, acl_content_3}.

\noindent\textbf{Propagation-Based Methods.} The spread of fake news on social media has garnered significant attention. Some studies model the propagation process as a sequence \citep{seq_model1, seq_2}, often using attention mechanisms to capture long-range dependencies \citep{propagation_attention_PLAN, propagation_attention_GCAN, seq_att}. Furthermore, fake news propagation is sometimes modeled as a tree, making it suitable for graph classification techniques \citep{propagation_GNN_duck_2022, DBLP:conf/acl/WeiHZYH20, dou2021rumor, acl_propagation_2, acl_propagation_3, propagation_attention_GCAN, acl_propagation_1}. Some studies explore the role of users in propagation \citep{propagation_GNN_upfd_2021, upsr}, while others focus on the structure of the propagation tree \citep{propagation_GNN_RAGCL}.

\subsection{Fake News Dataset} 
\noindent\textbf{Content-centric Datasets} initially include news metadata, textual content, and supervised labels, forming the basis for early fake news detection tasks \citep{dataset_textonly_NELA-GT-2018, dataset_textonly_realnews}. To reduce bias from semantic and lexical content, factual evidence and multi-modal data are integrated, enhancing the detection framework \citep{dataset_evi_HOVER, dataset_evi&multimodal_mr2, dataset_evi&multimodal_mocheg}.

\noindent\textbf{Propagation-based Datasets} capture key propagation data, such as retweets, comments, and likes, aiding fake news detection by analyzing propagation patterns \citep{fakenewsnet, twitter16, weibo16}. \citet{domain_2_MDFEND} constructs the large Chinese dataset Weibo-21, adding domain tags to the news metadata and incorporating propagation data. To address the limitations of single-source datasets, \citet{mcfend} develops the multi-source benchmark MCFEND for Chinese fake news detection, using data from social platforms, messaging apps, and traditional news.

However, previous works primarily focus on propagation patterns within a single platform. In contrast, we introduce MPPFND, a pioneering multi-platform dataset that offers richer social context, laying the foundation for deeper investigation into fake news propagation characteristics.

\section{Fake News Detection with Multi-Platform Propagation}
% \section{Empirical Study of Propagation across Multiple Social Platforms}
% \iffalse
% 在本节中我们将介绍我们MPPFND数据集的收集方法与MPPFND数据集的统计特征，我们获取数据集的整体流程图如图所示。此外我们还会给出我们数据集的统计特征，并给出我们对不同平台传播特征的分析。在表中我们给出了我们与其他数据集的对比，以展示我们与其他工作的不同。
% （对比表格）（流程图 fakenewsnet twtter1516）
% \fi
% \iffalse
% In this section, we describe the process of data collection for MPPFND dataset. The overall flowchart for obtaining the dataset is shown in the Figure~\ref{fig:data_crawling}. In addition, we will provide statistical features of our dataset and analyze the propagation characteristics of different platforms. In the Table, we present our comparison with other datasets to demonstrate our differences from other works.
% \fi

Multi-platform fake news detection aims to detect fake news using propagation structure information from multiple social media platforms. 
To learn the intricate structure and inherent characteristics between platforms, we first construct a fake news dataset encompassing the propagation structures across multiple platforms (short for \textbf{MPPFND}). Subsequently, we conduct a thorough empirical analysis based on this dataset to learn platform-specific and platform-shared propagation patterns for fake news detection.

\subsection{Dataset Construction}

In this subsection, we describe the process of data collection for our dataset.

\subsubsection{Data Collection}

% \iffalse
% 我们首先从两个事实核查网站（pol snop）收集了原始新闻claim，我们收集的话题包括COIVID-19和俄乌战争等。
% 对于每一个新闻，重要的信息包括其真假标签与claim的文本，除此以外我们还收集了核查员、核查时间和标题等信息。
% \fi
% \iffalse
% We first collected original news claims from two fact checking websites politifact\footnote{https://www.politifact.com/} and snopes\footnote{https://www.snopes.com/}, including topics such as COIVID-19 and the Russia Ukraine war.
% For each news, important information includes its true or fake labels and the content of the claim. In addition, we also collected information such as the verifier, verification time, and title 123
% \fi

% \iffalse
% (修改)
% 我们以获取到的新闻文本作为关键词，在不同的社交媒体平台搜索对应的帖文作为一级传播节点；并收集其对应的评论作为二级传播节点。最终的传播结构图如图所示。
% \fi

 We utilize fact-checking websites\footnote{https://www.politifact.com/}\footnote{https://www.snopes.com/} to obtain news contents and obtain the ground-truth label for fake news based on annotations provided by expert fact checkers. 
Then, to achieve multi-platform social engagements, we create search queries based on the headlines to retrieve posts on three popular social platforms (i.e., Youtube, X, Reddit). X is a short-text social media platform, which are featured by rapid information diffusion. Reddit are long-text social media platform that can contain rich information. YouTube, as a video-sharing platform, hosts discussions that often include richer video evidence. Additionally, we further fetch the user engagements including comments, reposts and likes towards these posts on the above social platforms. MPPFND encompasses 4965 data samples spanning two semantic topics and three platforms.

\subsubsection{Label Mapping Strategy}

% \iffalse
% 我们将每个传播节点增加其来自何种平台的特征。(对于新闻根节点，其来自两个不同的事实核查网站，我们分别以0,1标注。对于一级传播节点，来自三个不同的社交媒体平台，我们分别用2-4对不同的社交媒体平台进行标注。)对于二级传播节点，其属于从属的以及传播节点的平台，故此只需要用其对应的一级传播节点的特征标注即可。
% \fi

The fact-checking platforms provide a variety of labels. 
Following \citet{domain_2_MDFEND}, we convert this into a binary classification problem by using the following label mapping strategy. For PolitiFact, which has eight original labels in total, we designate \textit{mostly-true}, \textit{half-true}, and \textit{true} as \textit{True}, while the remaining labels are marked as \textit{False}. For Snopes, which has twelve original labels, we mark \textit{mostly-true} and \textit{true} as \textit{True}, and the rest as \textit{False}.

% \subsubsection{Data Statistics}
% \subsubsection{Preprocessing}

% We collect 4,965 news articles covering both the COVID-19 pandemic and the Russia-Ukraine conflict, comprising 3,700 fake news and 1,265 true news articles. 
% To balance the labels of the data, we randomly selected 1,233 pieces of news with fake labels from the 3700 pieces and combine them with 1265 pieces of true news to form the final dataset. 

% \subsection{Analysis of Multi-Platform Propagation for Fake News Detection}
\subsection{Analysis of Propagation Properties across Multiple Social Platforms}

To explore the differences across different social media platforms, we conduct in-depth analyses based on our MPPFND dataset.

\begin{table}
  \centering
  \resizebox{0.48\textwidth}{!}{
  \begin{tabular}{lccc}
    \hline
    \textbf{Domain}        & \textbf{\#Total Nodes} & \textbf{Tree-Width} & \textbf{\#Avg. Nodes} \\
    \hline
    Youtube                & 450,711                  & 774                    & 273.99                         \\
    Youtube (Fake)          & 404,348                  & 774                    & 304.48                         \\
    Youtube (True)          & 46,363                   & 600                    & 146.26                         \\
    \hline
    X                  & 11,059                   & 96                     & 6.00                           \\
    X (Fake)            & 9,458                    & 96                     & 6.50                           \\
    X (True)            & 1,601                    & 55                     & 4.13                           \\
    \hline
    Reddit                 & 35,258                   & 300                    & 323.47                         \\
    Reddit (Fake)           & 29,297                   & 300                    & 321.95                         \\
    Reddit (True)           & 5,961                    & 200                    & 331.17                         \\
    \hline
  \end{tabular}
  }
  \caption{
    Summary of data for various domains including total nodes, tree-width, and average nodes per graph.
  }
  \label{tab:data-summary}
\end{table}

\begin{figure}[t]
  \centering
  % 第三张图：comment_similarity（原图3，现在放在左上角）
  \begin{subfigure}[b]{0.46\columnwidth}
    \includegraphics[width=\textwidth]{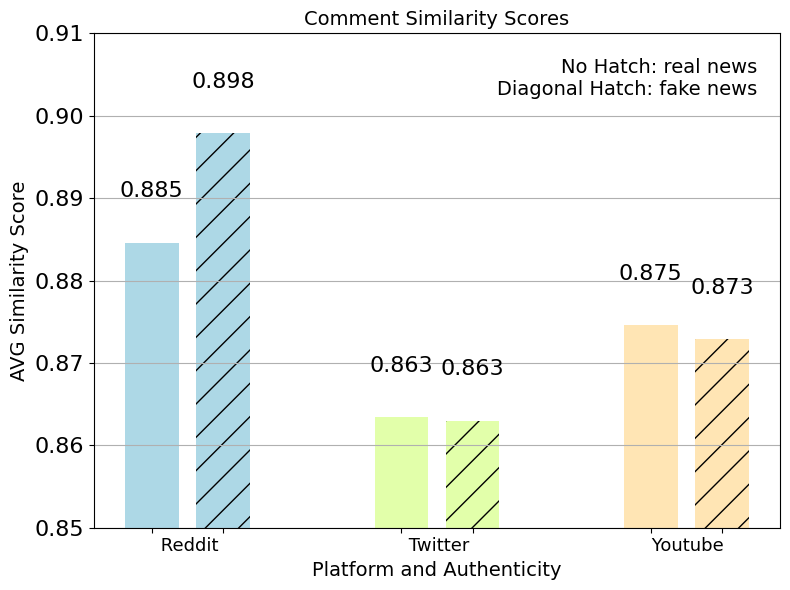}
    \caption{Comment similarity with the claim of each platform in our dataset.}
    \label{fig:comment_similarity}
  \end{subfigure}
  \hfill
  % 第二张图：comment_length（原图2，保持不变）
  \begin{subfigure}[b]{0.46\columnwidth}
    \includegraphics[width=\textwidth]{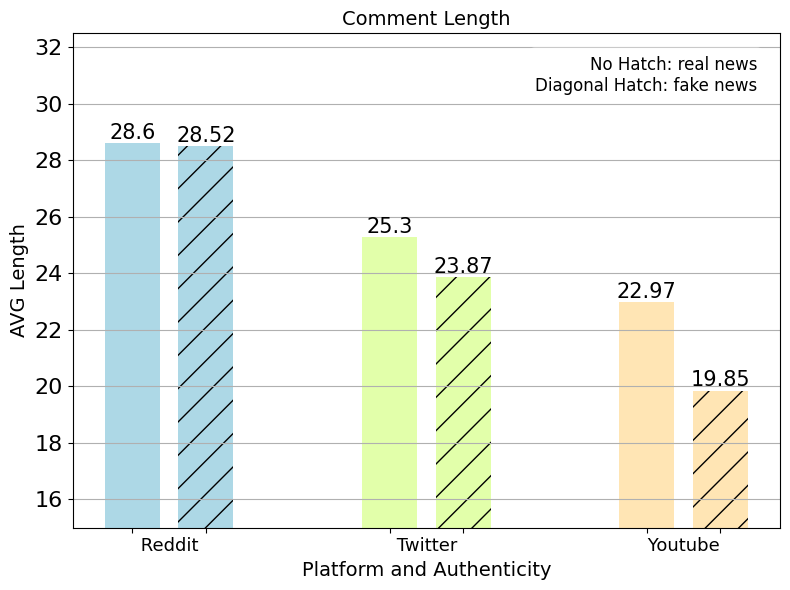}
    \caption{Analysis of comment length on each social platform in our dataset.}
    \label{fig:comment_length}
  \end{subfigure}
  
  \vspace{\floatsep} % 垂直间距
  
  % 第一张图：news_duplication（原图1，现在放在左下角）
  \begin{subfigure}[b]{0.46\columnwidth}
    \includegraphics[width=\textwidth]{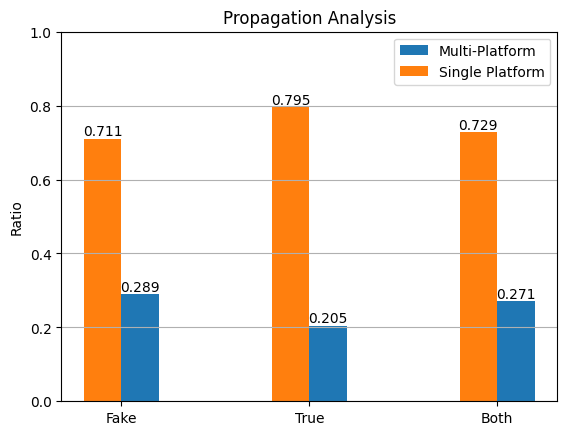}
    \caption{The proportion of news claim across multiple or on a specific platform.}
    \label{fig:news_duplication}
  \end{subfigure}
  \hfill
    \begin{subfigure}[b]{0.46\columnwidth}
    \includegraphics[width=\textwidth]{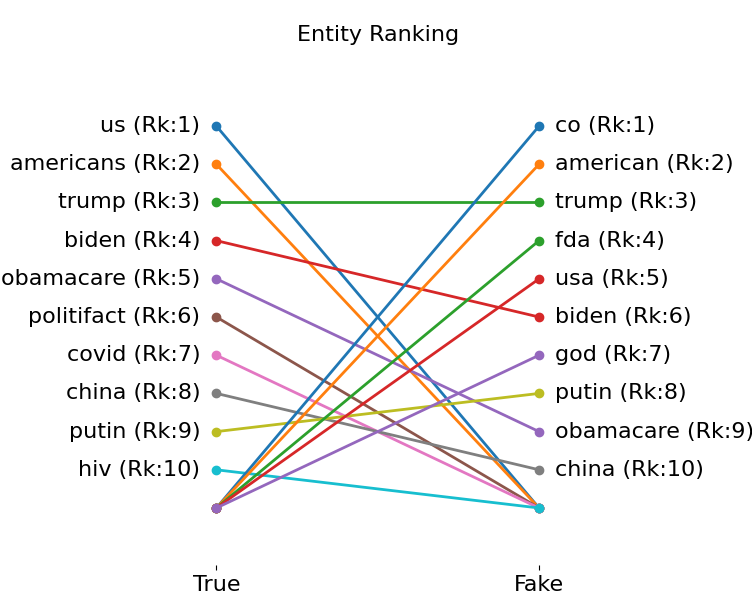}
    \caption{The Top 10 entities most frequently appearing in news and its comments.}
    \label{fig:ea}
  \end{subfigure}
  % \begin{subfigure}[b]{0.45\columnwidth}
  %   \includegraphics[width=\textwidth]{figure/te.png}
  %   \caption{Entity analysis of comments on Twitter.}
  %   \label{fig:te}
  % \end{subfigure}
  
  % \vspace{\floatsep} % 垂直间距

  % \begin{subfigure}[b]{0.4\columnwidth}
  %   \includegraphics[width=\textwidth]{figure/ye.png}
  %   \caption{Entity analysis of comments on Youtube.}
  %   \label{fig:ye}
  % \end{subfigure}
  % \hfill
  % \begin{subfigure}[b]{0.4\columnwidth}
  %   \includegraphics[width=\textwidth]{figure/re.png}
  %   \caption{Entity analysis of comments on Reddit.}
  %   \label{fig:re}
  % \end{subfigure}
  % \hfill
  
  \caption{ (a) Comment similarity, (b) Comment length, (c) News duplication, (d) Entity analysis.}
  \label{fig:combined_figures}
\end{figure}

% \begin{figure}[t]
%   \centering
%   \includegraphics[width=0.86\columnwidth]{figure/news_duplication.png}
%   \caption{The proportion of news claim across multiple platforms and on a specific platform. 
%   % Blue color represents number of news spread across multiple platforms, orange indicates number of news spread on a single platform.
%   }
%   \label{fig:news_duplication}
% \end{figure}

% \begin{figure}[t]
%   \centering
%   \includegraphics[width=0.86\columnwidth]{figure/comment_length.png}
%   \caption{Analysis of comment length on each social platform in our dataset.}
%   \label{fig:comment_length}
% \end{figure}

\begin{figure}[t]
  \centering
  \begin{subfigure}[b]{0.21\textwidth}
    \centering
    \includegraphics[width=\linewidth]{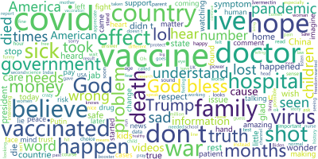}
    \caption{Comments on YouTube}
    \label{fig:cloudyou2}
  \end{subfigure}
  \hfill
    \begin{subfigure}[b]{0.21\textwidth}
    \centering
    \includegraphics[width=\linewidth]{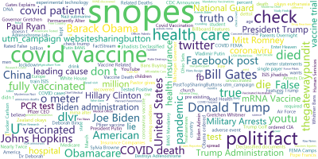}
    \caption{Comments on YouTube}
    \label{fig:cloudyou1}
  \end{subfigure}
  \hfill
  
  \begin{subfigure}[b]{0.21\textwidth}
    \centering
    \includegraphics[width=\linewidth]{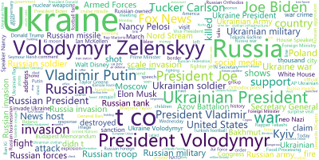}
    \caption{Comments on Reddit}
    \label{fig:cloudre2}
  \end{subfigure}
  \hfill
  \begin{subfigure}[b]{0.21\textwidth}
    \centering
    \includegraphics[width=\linewidth]{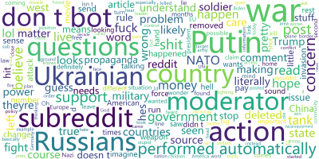}
    \caption{Comments on Reddit}
    \label{fig:cloudre1}
  \end{subfigure}
  \hfill
  
  \begin{subfigure}[b]{0.21\textwidth}
    \centering
    \includegraphics[width=\linewidth]{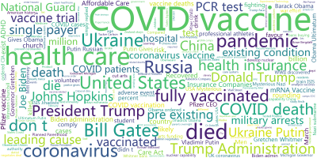}
    \caption{Comments on X}
    \label{fig:cloudtw2}
  \end{subfigure}
  \hfill
  \begin{subfigure}[b]{0.21\textwidth}
    \centering
    \includegraphics[width=\linewidth]{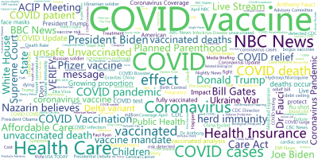}
    \caption{Comments on X}
    \label{fig:cloudtw1}
  \end{subfigure}
  
  \caption{Word cloud of comments in True and Fake claims. Left: Fake claim comments on YouTube, Reddit, and X. Right: True claim comments on YouTube, Reddit, and X.}
  \label{fig:combined_clouds}
\end{figure}

\subsubsection{Analysis of Propagation Statistics}

To investigate propagation structures across platforms, we analyze the proportion of news claims on multiple social media platforms, as shown in Table~\ref{tab:data-summary}.
\textbf{The spread of fake and true information shows distinct platform-specific and common traits.} For example, fake news engagement is higher on YouTube and X but lower on Reddit. Fake news also spreads across more platforms than true news, indicating broader reach.
\textbf{Skewed distributions}, such as fewer nodes on X, highlight potential propagation bias in low-resource platforms for multi-platform fake news detection.
Figure~\ref{fig:news_duplication} shows the distribution of news across single and multiple platforms.
\textbf{Incomplete cross-platform propagation data} underscores the need for adaptive detection models.
\textbf{True news is more common on single platforms, while fake news spreads more across multiple platforms.} This suggests fake news has stronger cross-platform diffusion capability, likely due to its sensational or controversial nature, prompting more frequent sharing.

\subsubsection{Analysis of Comment Style}
To further explore comment styles across different platforms, we first analyze the length of comments to determine whether users have fully expressed their opinions. As shown in Figure~\ref{fig:comment_length}, we observe that: 1) On X and YouTube, comments on fake claims are shorter than true claims, whereas this phenomenon is not observed on Reddit. This reveals that \textbf{discussions on Reddit are more extensive and comprehensive than on other platforms}. 2) \textbf{Not all comments across platforms contain sufficient discriminative features for detecting fake news.} Therefore, it is necessary to use different models to extract the features of comments from each platform.

Figure~\ref{fig:combined_clouds} illustrates the differences in keywords used across various platforms. 
Specifically,  X emphasizes short texts and immediacy, hosting real-time discussions and sharing current events more than Reddit and YouTube. In contrast, Reddit and YouTube tend to feature more fragmented content, focusing on hot topics and current affairs.
We also observe that both Reddit and X feature discussions on COVID-19 vaccines and political figures (e.g., Biden and Trump). However, Reddit covers more topics related to the Ukraine conflict, Russia, and related actions. Both YouTube and X discuss COVID-19 vaccines, but YouTube places more emphasis on topics related to doctors, countries, and beliefs.
Through entity analysis in Figure~\ref{fig:ea}, we found that \textbf{comments on true claims, in contrast to those on fake claims, often incorporate more detailed evidence and positive reinforcement}, including references to prominent figures (e.g., Trump, Putin, Biden) or affirming terms (e.g., support, truth, believe). This highlights a clear distinction between the nature of comments on true versus fake claims. Notably, this trend is consistent across all three platforms, pointing to shared patterns in how information spreads.

\subsubsection{Analysis of Comment Emotion and Echo Chambers}

\begin{figure}[t]
    \centering
    \begin{subfigure}[b]{0.28\textwidth}
        \centering
        \includegraphics[width=\textwidth]{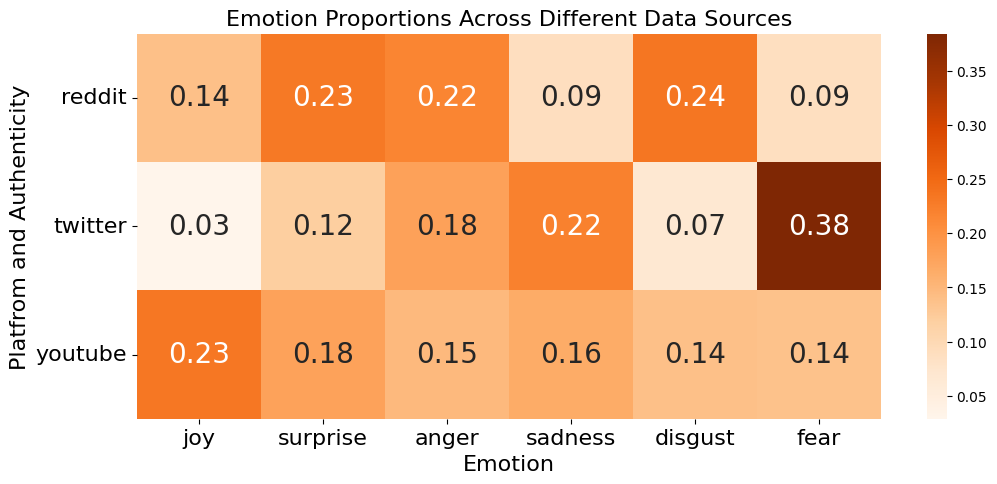}
        \caption{Comment emotion in fake claim.}
        \label{fig:emotion_fake}
    \end{subfigure}
    \begin{subfigure}[b]{0.28\textwidth}
        \centering
        \includegraphics[width=\textwidth]{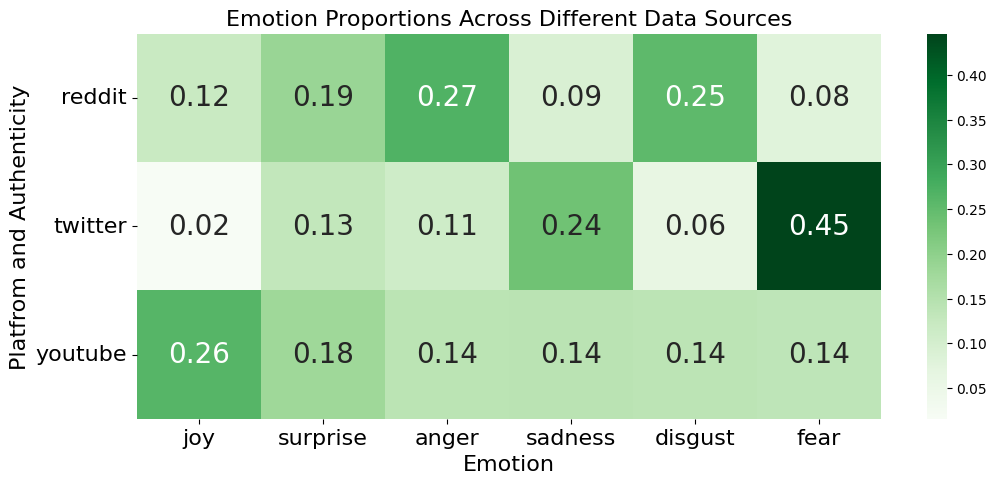}
        \caption{Comment emotion in true claim.}
        \label{fig:emotion_true}
    \end{subfigure}
    \hfill
    \caption{Emotion feature of each social platform comment in news claim. Due to the prevalence of neutral emotions across all platforms, we omit neutral emotions to highlight differences. }
    \label{fig:emotion}
\end{figure}

To explore the emotional tendencies of comments and quantitatively describe the platform's Echo Chambers \cite{echo}, we extract the emotional features of comments shown in Figure~\ref{fig:emotion}. We find that \textbf{emotional features on YouTube are difficult to use in distinguishing the credibility of news.} On Reddit and X, there is a significant difference in the proportion of comments expressing different emotions between fake news and real news. However, on YouTube, this difference is almost negligible.

The relevance of comments to news claims is shown in Figure~\ref{fig:comment_similarity}. We find that \textbf{Reddit exhibits a stronger Echo Chamber effect.} Comments on Reddit are highly relevant to the claims themselves, followed by YouTube, whereas comments on X show weaker relevance. On Reddit, comments on fake claims are more relevant than comments on true claims; on YouTube, the reverse is observed. This indicates that the Echo Chamber effect on Reddit is more pronounced compared to the other two platforms.

% \subsubsection{Conclusion of the Analysis}
% Based on the above results user engagement and comment analysis, we draw the following conclusions:
% \begin{itemize}
%     \item 
%     Reddit users prefer discussing true news, whereas X and YouTube platforms exhibit the opposite preference, suggesting that more reliable discussions can be obtained from Reddit comments.
%     \item 
%     Comments on X and YouTube are more varied, whereas Reddit exhibits a stronger Echo Chamber Effect. 
%     \item 
%     We can find that news claim with higher user engagement on YouTube and X are more likely to be fake, whereas this characteristic is not evident on Reddit. The method of using comment sentiment to differentiate news authenticity can be applied on X and Reddit, whereas it is not effective for YouTube. Thus, it is evident that differential modeling across various platforms is necessary for detecting fake news.
% \end{itemize}

\section{Adaptive Propagation Structure Learning Network}
% \section{Fake News Detection using rich social context on Multi-Platform Propagation}

% \iffalse
% 在本节中我们将详细介绍我们提出的多平台自适应模型的框架。整体结构如图所示，可以看出我们的模型有三个重要的组件。我们首先利用bert模型对新闻的文本特征进行提取，获取新闻的表示特征。然后我们对每个节点的文本使用另一个bert模型提取节点特征，并利用多个GNN网络获取不同的平台特征传播特征，并通过注意力机制拼合不同社交媒体平台的特征向量。最终使用一个多层感知器汇总所有的特征并用sigmoid函数使其输出新闻的真假概率，使用交叉熵损失训练函数。
% \fi

\begin{figure}[t]
  \centering
  \includegraphics[width=\linewidth]{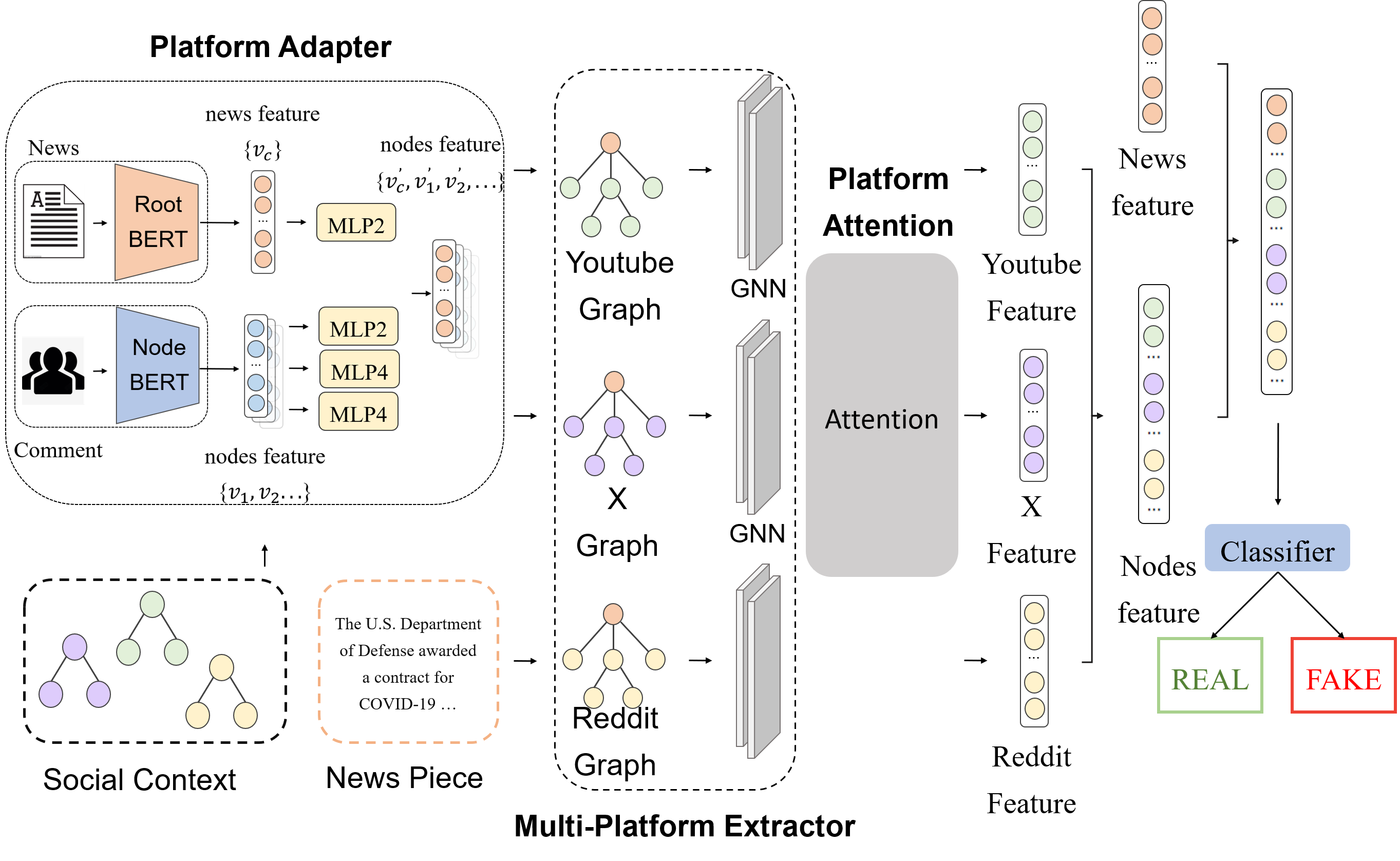}
  \caption{Overview of the proposed framework APSL.}
  \label{fig:model}
\end{figure}

In this section, we propose an efficient Adaptive Propagation Structure Learning Network (APSL) for multi-platform fake news detection. As shown in Figure~\ref{fig:model}, we use a text feature encoder to extract textual features from news articles and comments. To capture platform-specific features, we design a {Platform Adapter} for different nodes. Given the varying impact of propagation characteristics across platforms, we employ a {multi-platform propagation encoder} to extract propagation structure features. To focus on relevant comments, we use a {feature fusion module} to integrate platform-specific textual and news claim features. Finally, a multilayer perceptron (MLP) consolidates all features, and a Sigmoid function outputs the news authenticity probability, with the model trained using cross-entropy loss.

% \subsection{Problem Formulation}

\textbf{Multi-platform fake news detection} aims to detect fake news using propagation structure information from multiple social media platforms. 
Formally, given a labeled claim dataset as $ D = \{ (c, s, y) \} $, where $ c $ represents the news claim and \( y \) represents the comments on the news claim, and $ s $ contains the comment of the news claim. $ s = \{ s_{1}, s_{2}, \ldots, s_{m} \} $, where $ m $ indicates the number of comments. The model aims to predict the True or fake label of the news claim.

\subsection{Text Encoder}

% \iffalse
% Specifically, we denote a labeled claim dataset as C = { ($c^i$, $s^i$) }

% 我们使用BERT模型提取文本特征。考虑到新闻文本本身对其真假标签有重要影响，而评论信息的噪声更大，我们使用单独的根节点bert模型提取claim的特征。我们训练一个另外的bert模型来处理除根结点外的其他节点信息，以此减小传播结构中评论的噪音对模型建模新闻文本特征的影响。

% \fi

% Denote a labeled claim dataset as $ D = \{ (c^1, s^1, y^1), (c^2, s^2, y^2), \ldots, (c^n, s^n, y^n) \} $, where $ c^i $ represents the i-th news claim and \( y^i \) represents the label of the news claim. $ s^i $ contains the comment of the i-th news claim. $ s^i = \{ s_{1}^i, s_{2}^i, \ldots, s_{m_{i}}^i \} $, where $ m_{i} $ indicates the number of comments.

% For each data $ (c, \{ s_{1}, s_{2}, \ldots, s_{m_{i}} \}, y) $, since the news content itself significantly impacts its authenticity label, whereas comment information tends to be noisier, we employ a separate Text Encoder to extract features for the root node of the claim. Additionally, we train another Text Encoder to handle information from nodes other than the root node. This approach mitigates the impact of noisy comment data on the model's ability to capture features from the news content.

We utilize two text encoders to extract semantic features from the claim and its propagation on social platforms.
Given the news content of the claim $c$ and the $j$-th comment $s^k_{j}$ on platform $k$, we use the text encoder to extract text embedding of claim and comments as $\mathbf{c}$ and $\mathbf{s}_{j}^k$.

Considering the significant differences in comments across multiple platforms, we introduce a learnable vector $\mathbf{p}^{k}$ to adaptively capture specific semantic features on platform $k$. 

The platform-adaptive representations can be computed as follows:

\begin{equation}
  \label{eq:node feature2}
  \mathbf{\tilde{s}}_{j}^{k} = \text{Softmax}(\mathbf{W}^{k}(\mathbf{p}^{k} \cdot \mathbf{s}^k_{j})+\mathbf{b}^k)
\end{equation}
where $\mathbf{W}^{k}$ and $\mathbf{b}^k$ are trainable parameters for platform $k$.

\subsection{Multi-Platform Propagation Encoder}

% \iffalse
% 对于节点特征\( \{ v_{c}, v_{1}, v_{2}, \ldots, v_{m_{i}} \} \)，我们首先根据每个节点所属的平台对不同的节点特征做一次数据增强。具体而言，我们为每个平台设置了一个可学习的向量e作为平台的特征向量。再用平台向量经过一个线性层，并以该输出作为一个节点特征的权重。通过将该输出作为节点的attention来获取获得不同的平台相应的强化节点特征表示。之后，将经过平台传播强化的节点特征作为GNN的初始节点特征并获取其传播结构信息。
% 此外对于不同的平台，我们分别训练一个GNN来获取其对应的平台传播特征表示，以此解决不同平台传播特征不同的问题。
% 得到不同平台的特征后，我们使用attention机制来合并平台特征。具体而言，我们以news claim的文本特征\(  v_{c}^i \)作为Q，以不同平台的特征\(  \{ v_{g}^1, v_{g}^2, \ldots, v_{g}^k \} \)作为P和V，以此提取与新闻内容相关的传播特征信息。最后将得到的结果拼接得到最终的传播特征\( v_{s} \)
% \fi

\paragraph{Graph Construction}For a news claim, there are multiple propagation graphs across different platforms.For platform $k$, the propagation graph are denoted as $G^k =  \{ V^k, E^k \}$. $V^k = \{c, s_1^k, s_2^k, ..., s_M^k\}$ denotes the set of nodes including the claim $c$ and its propagation $s_1^k,  ..., s_{M_k}^k$ on the platform $k$. $M_k$ is the number of comments on the platform.
Their initial node embeddings can be computed with $\mathbf{c}$ and $\mathbf{\tilde{s}}_i^k$. $ E^k = \{ e_{1}, e_{2}, \ldots, e_{n_{e}} \} $ denotes the set of edges, where the edges between nodes are built based on the relationships between reposts and comments.

\paragraph{Learning Platform-specific Propagation Structures}In addition, for different platforms, we train a separate GNN for each platform to obtain the corresponding propagation feature representations, which allow us to learn latent structural features from the propagation specific to each platform. We use global add pooling to obtain $ \mathbf{h_{g}} $.
% \begin{equation}
%   \label{eq:platform feature}
%   \mathbf{h}_{g}^k = \text{GNN}^{k}({G^k})
% \end{equation}

\paragraph{Multi-Platform Propagation Fusion Module}

Considering that some social context features may be irrelevant to the news claim, we use an attention mechanism to select the portions that are pertinent to the news content. Specifically, after obtaining features from different platforms, we use the textual features of the news claim $\mathbf{c}$ to guide the fusion of multi-platform propagation features $\{ \mathbf{h}^1, \mathbf{h}^2, \ldots, \mathbf{h}^{|P|} \}$. The mechanism then extracts the propagation features that are most relevant to the news content. Finally, we concatenate them to form the final propagation feature $\mathbf{h_{s}}$.
\begin{equation}
  \label{eq:attention platform feature1}
    \mathbf{h}^k=\text{Softmax}(\frac{\mathbf{c} {(\mathbf{h}_{g}^{k})}^T}{\sqrt{d_{h_{g}^k}}})\mathbf{h}_{g}^k
    % \text{Attention}(\mathbf{ v_{c},  h_{g}, h_{g} })=
\end{equation}
% \begin{equation}
%   \label{eq:attention platform feature2}
%   \mathbf{h_{s}} = \text{Concat}(\mathbf{h}^1, \mathbf{h}^2, \ldots , \mathbf{h}^{|P|})
% \end{equation}
% where $P$ is the set of multiple social platforms, and $|P|$
%  is the number of the platform set.
 % \{Twitter, Reddit, Youtube\}

% \iffalse
% \begin{equation}
%   \label{eq:con}
%   \mathbf{L_{PCL}} = -\frac{1}{\mathbf{|B|}} \sum\limits_{i=1}^{\mathbf{|B|}} \sum\limits_{j=1}^{\mathbf{|B|}}\mathbf{l}_{y^i=y^j}
%   \log_{}{
%   \frac{exp(sim(\mathbf{h}_i^P, \mathbf{h}_j^P)/\tau)}{
%   \sum\nolimits_{m=1}^{\mathbf{|B|}}exp(sim(\mathbf{h}_m^P, \mathbf{h}_m^P)/\tau)
%   }
%   }
% \end{equation}
% \fi

\subsection{Fake News Detector}

% \iffalse
% 使用传播节点bert模型提取传播节点的特征
% 再利用GCN获取高层传播图特征，并取图的全局特征向量作为新闻传播结构的特征
% 在上述两个小节中我们分别获取了新闻的文本特征信息与平台的传播结构信息，我们按权重将这两个特征进行融合，如公式所示,其中beta和alpha为超参数。
% 在得到融合的特征后，我们使用一个具有softmax输出的MLP对新闻的真假标签进行预测。该模型使用二元交叉熵损失函数进行训练，并使用 SGD 进行更新。
% \fi

We concatenate the claim representation $\mathbf{c}$ and the propagation representations to fuse them into $\mathbf{h}_{s}$.

Based on the fusion representations, we use an MLP with Sigmoid output to predict the true or fake labels of the news:
\begin{equation}
  \label{eq:predict}
  \hat{y} = \text{Sigmoid}(\mathbf{W}^{d}(\text{Concat}(\mathbf{c}, \mathbf{h}_s)+\mathbf{b}^d)
\end{equation}
where $\mathbf{W}^{d}$ and $\mathbf{b}^d$ are learnable parameters.

\subsection{Platform-aware Contrastive Learning}

Due to the variability and noise in the comments, we aim to capture the general features of each platform. Inspired by the success of self-supervised contrastive learning \citep{con_1, con_2}, we use contrastive learning to automatically learn the general propagation features. The objective function is computed as:
\begin{equation}
  \label{eq:con}
  \resizebox{0.48\textwidth}{!}{$
  \mathcal{L}_{PCL} = -\sum\limits_{k\in\mathbf{P}}^{} \frac{1}{|B|} \sum\limits_{i=1}^{|B|} \sum\limits_{j=1}^{|B|}\mathbf{l}_{y^i=y^j}  
  \log_{}{
  \frac{exp(sim(\mathbf{h}_i^k, \mathbf{h}_j^k)/\tau)}{
  \sum\nolimits_{m=1}^{|B|}exp(sim(\mathbf{h}_m^k, \mathbf{h}_m^k)/\tau)
  }
  }$}
\end{equation}
where $|B|$ is the batchsize, $\mathbf{l}$ is an indicator. $k$ indicates a specific platform and $\tau$ controls the temperature

This model is trained using a binary cross entropy loss function for fake news detection denoted as $\mathcal{L}_{pred}$. We combine all of the loss functions together to jointly train our model, and can be represented as follows:
\begin{equation}
  \label{eq:bce}
   \mathcal{L}_{final} = \mathcal{L}_{pred} + \gamma\mathcal{L}_{PCL}
\end{equation}
where $\gamma$ is hyperparameter that controls the balance between the two loss components.

\section{Experiments}

% \iffalse
% 数据预处理：
% 加段话

% 数据分析：
% 情感分析（找工具），评论和claim本身的相似度的分析
% 实体分析

% 实验：
% 文本虚假信息检测模型 bigcn模型 案例分析 困惑矩阵分析（不同平台传播数据对虚假新闻标签的预测变化）

% 消融：
% 模块消融 、 平台消融

% 在本节中，我们将详细介绍我们的实验设置，并将我们提出的MPPFND模型与其他的基线模型在我们的数据集上进行了实验，以此验证我们提出的模型的有效性。
% \fi

% In this section, we provide a detailed introduction to our experimental setup and conduct experiments on our dataset with our propose model and other baseline models to verify the effectiveness of our proposed model.
\subsection{Experimental Setups}
\paragraph{Baselines}

% \iffalse
% 我们使用了三组不同的虚假新闻检测模型与我们提出的APSL的效果进行比较：
% 仅使用文本特征的虚假新闻检测模型：BERT以及RoBERTa
% 使用文本和使用序列模型建模传播结构信息的虚假新闻检测模型：BERT + GRU和BERT+ Transformer。在实验中我们使用BERT模型来建模新闻文本特征，使用序列模型对传播结构特征进行建模，最终融合特征得到结果。
% 使用文本和使用GNN传播结构信息的虚假新闻检测模型：BERT + GNN ，其中我们使用的GNN模型包括GCN、GAT和SAGE。在实验中我们使用BERT模型来建模新闻文本特征，使用GNN模型对传播结构特征进行建模，最终融合特征得到结果。
% 将以上方法与我们提出的自适应多平台传播模型进行对比，以此验证对不同平台的传播信息进行建模是否对虚假信息检测任务的效果有提升。
% \fi

We compare with content-based and propagation-based fake news detection models. For content-based models, we used BERT \citep{bert}, RoBERTa \citep{RoBerta}, EANN \cite{eann} and gpt-4o-mini. The sequence model we use including GRU \citep{gru} and Transformer \citep{transformer}. The fake news detection model we use including  BiGCN \citep{bigcn}, UPFD \citep{propagation_GNN_upfd_2021} and UPSR \citep{upsr}. 
% Compare the above methods with our proposed model to verify whether modeling the propagation information of different platforms improves the effectiveness of fake news detection tasks.

\paragraph{Implementations}

% \iffalse
% 实验设置
% 我们使用pytorch框架来实现我们的模型。考虑到数据集中真假标签的不平衡性，我们对标签为负的样本做采样以确保数据的均衡性。我们将数据以7：1：2的比例随机分为训练集/验证集/测试集，并确保每个数据集中拥有传播结构的claim与无传播结构的claim比例也保持均衡。bert的单词嵌入向量的维数固定为768，最大输入长度设置为64。我们对所有模型使用统一的图嵌入维数（768），批量大小（8），优化器 （Adam）。对于建模传播结构的模型，\mathbf{\alpha}和\mathbf{\beta}均被默认设为1和0.8.针对我们提出的APSL模型，我们选择了SAGE作为Platform Extractor，RoBERTa作为Text Feature Extractor。

% \fi

 Considering the imbalance between true and fake labels in the dataset, we sample news with negative labels to ensure data balance. We randomly divide the data into training, validation, and testing sets in a 7:1:2 ratio. The dimension of word embedding vector is fixed at 768. The temperature $\tau$ is 0.1, and hyperparameter $\gamma$ is 0.3.We train the model with Adam optimizer. We conduct three experiments and report the average of the results.

\subsection{Main Results}

% \iffalse
% 经过实验，我们认为对不同的社交媒体平台的传播结构差异进行分析与建模，所提取的特征有助于对新闻文本的真实性分类提供帮助。
% (1)当平台的传播数据较少时，用GNN建模的传播结构数据无法得到充分的训练，导致结果相较于文本检测模型更低。

% 我们将我们的APSL模型与文本检测方法、序列模型建模社交上下文的方法和用图神经网络建模社交上下文方法进行了比较，以证明其有效性。结果如表所示。实验结果进一步揭示了一些有见地的观察结果。
% (1)当平台的传播数据较少时，用GNN建模的传播结构数据无法得到充分的训练，导致结果相较于文本检测模型更低。
% (2)增加了对社交上下文建模的模型总体上比仅对文本建模的模型的效果更好，这表明社交上下文的数据非常重要。
% (3)与其他模型相比，我们的APSL模型总体上表现更好，这说明对不同平台的评论进行差异化建模对虚假新闻的检测是有用和必要的。

% \fi

\begin{table}
  \centering
  \resizebox{0.45\textwidth}{!}{
  \begin{tabular}{lcccc}
    \hline
    \textbf{Model}        & \textbf{Accuracy} & \textbf{Precision} & \textbf{Recall} & \textbf{F1} \\
    \hline
    EANN                  & 0.6367            & 0.6378             & 0.6359          & 0.6351 \\
    GPT-4o                  & 0.6946            & 0.6946             & 0.6944          & 0.6944 \\
    \hline 
    \multicolumn{4}{l}{\textit{Backbone=BERT}}
    \\ \hline 
    Bert                  & 0.6707            & 0.6841             & 0.6687          & 0.6629 \\
    % \hline
    Transformer      & 0.6766            & 0.6829             & 0.6752          & 0.6727 \\
    GRU              & 0.6467            & 0.6498             & 0.6456          & 0.6437 \\
    % \hline
    UPFD w/ GCN              & 0.6766            & 0.6819             & 0.6754          & 0.6733 \\
    UPFD w/ GAT              & 0.6766            & 0.6835             & 0.6752          & 0.6724 \\
    UPFD w/ SAGE             & 0.6846            & 0.6887             & 0.6835          & 0.6821 \\
    BiGCN            & 0.6766            & 0.6783             & 0.6759          & 0.6752 \\
    UPSR                & 0.6786            & 0.7013             & 0.6762          & 0.6674 \\
    % \hline
    \textbf{APSL}(Ours)                  & \textbf{0.7046}            & \textbf{0.7045}             & \textbf{0.7045}          & \textbf{0.7045} \\
    \hline
    \multicolumn{4}{l}{\textit{Backbone=RoBERTa}}
    \\ \hline
    RoBERTa               & 0.6507            & 0.6595             & 0.6489          & 0.6441 \\
    Transformer        & 0.6607            & 0.6692             & 0.6590          & 0.6549 \\
    GRU                & 0.6547            & 0.6773             & 0.6521          & 0.6410 \\
    % \hline
    UPFD w/ GCN                & 0.6427            & 0.6473             & 0.6413          & 0.6385 \\
    % \hline
    UPFD w/ GAT                & 0.6587            & 0.6745             & 0.6565          & 0.6488 \\
    UPFD w/ SAGE               & 0.6727            & 0.6730             & 0.6729          & 0.6726 \\
    BiGCN                & 0.6453            & 0.6436             & 0.6448          & 0.6436 \\
    UPSR                & 0.6627            & 0.6663             & 0.6615          & 0.6598 \\
    \textbf{APSL}(Ours)               & \textbf{0.6846}            & \textbf{0.6854}   & \textbf{0.6850}          & \textbf{0.6845} \\
    \hline
  \end{tabular}
  }
  \caption{Multi-platform fake news detection performance in our dataset.}
  \label{tab:model-evaluation}
\end{table}

The results for fake news detection on the MPPFND dataset with multi-platform propagation are shown in Table~\ref{tab:model-evaluation}. From the table, we have the following observations: 1) Propagation-based methods generally outperform content-based ones, highlighting the importance of propagation in fake news detection. 2) Some propagation-based methods, like the BiGCN model, incorporate multi-platform propagation but perform worse, possibly due to insufficient propagation in the MPPFND dataset. UPSR improves the modeling of incomplete propagation, outperforming BiGCN. Our APSL, by leveraging platform differences, models complex platform propagation more effectively, leading to better performance than existing methods. 3) Our model generally outperforms others, suggesting that differentiated modeling of comments across platforms is both useful and necessary for fake news detection.

\subsection{Ablation Study}

% \iffalse
% 在消融实验中，我们主要考虑我们设计的两个模块对模型效果的影响：一个是多平台的处理模块，我们将其变为单一平台的数据来测试多平台数据对实验的影响。另一个模块是基于平台的节点特征增强，我们尝试了不使用节点增强策略的模型。而当我们同时删去这两个模块时模型将退化为bert+GNN模型，该实验的结果在前文中已经给出，故不重复说明。
% 实验结果如表所示。从表中我们可以看出，仅使用单一平台训练的模型尤其在传播数据大量缺失时（这种情况存在于twitter和reddit平台）检测效果甚至不如仅使用新闻文本的模型。即使是在传播结构相对丰富的youtube平台，单平台的检测效果也逊色于多平台。此外当我们不使用节点增强策略时检测效果也有一定程度的下降。
% \fi

\begin{table}
    \centering
    \resizebox{0.48\textwidth}{!}{
        \begin{tabular}{lcccc}
        \hline
        \textbf{Method} & \textbf{Accuracy} & \textbf{Precision} & \textbf{Recall} & \textbf{F1} \\
        \hline
        % \textbf{APSL(Our Model)} & all & \textbf{0.6986} & \textbf{0.7091} & \textbf{0.6970} & \textbf{0.6936} \\
        % \hline
        
        % \multicolumn{6}{l}{\textit{Platform}} \\
        % \hline
        % \multirow{3}{*}{\shortstack{Single-Platform}} & X   & 0.6427 & 0.6444 & 0.6433 & 0.6423 \\
        % & Reddit    & 0.6547 & 0.6546 & 0.6545 & 0.6545 \\
        % & Youtube   & 0.6766 & 0.6783 & 0.6759 & 0.6752 \\
        % \hline
        % \multirow{3}{*}{\shortstack{Double-Platform}} & w/o X   & 0.6846 & 0.6882 & 0.6836 & 0.6823 \\
        % & w/o Reddit    & 0.6846 & 0.6874 & 0.6837 & 0.6827 \\
        % & w/o Youtube   & 0.6687 & 0.6806 & 0.6668 & 0.6615 \\
        % \hline
        
        % \multicolumn{6}{l}{\textit{Model}} \\
        \hline
        \textbf{APSL} & \textbf{0.7046} & \textbf{0.7045} & \textbf{0.7045} & \textbf{0.7045} \\
        % \hline
        w/o Platform Contrastive Learning & 0.6986 & 0.7091 & 0.6970 & 0.6936 \\
        w/o Node Enhancement & 0.6846 & 0.6874 & 0.6837 & 0.6827 \\
        w/o Platform Attention & 0.6727 & 0.6730 & 0.6729 & 0.6726 \\
        \hline
        
        \end{tabular}
    }
    \caption{Results of Ablation study results on the MPPFND dataset.}
    \label{tab:Ablation Study}
\end{table}

\begin{table}
    \centering
    \resizebox{0.48\textwidth}{!}{
        \begin{tabular}{lccccc}
        \hline
        \textbf{Dataset} & \textbf{Platform} & \textbf{Accuracy} & \textbf{Precision} & \textbf{Recall} & \textbf{F1} \\
        \hline
        \textbf{APSL (Our Model)} & all & \textbf{0.7046} & \textbf{0.7045} & \textbf{0.7045} & \textbf{0.7045} \\
        % \hline
        w/o Propagation &  & 0.6707 & 0.6841 & 0.6687 & 0.6629 \\
        \hline
        \multirow{3}{*}{\shortstack{Single-Platform}} & X   & 0.6427 & 0.6444 & 0.6433 & 0.6423 \\
        & Reddit    & 0.6547 & 0.6546 & 0.6545 & 0.6545 \\
        & Youtube   & 0.6766 & 0.6783 & 0.6759 & 0.6752 \\
        \hline
        \multirow{3}{*}{\shortstack{Double-Platform}} & w/o X   & 0.6846 & 0.6882 & 0.6836 & 0.6823 \\
        & w/o Reddit    & 0.6846 & 0.6874 & 0.6837 & 0.6827 \\
        & w/o Youtube   & 0.6687 & 0.6806 & 0.6668 & 0.6615 \\
        \hline
        
        \end{tabular}
    }
    \caption{Results of using propagation across different social platforms.}
    \label{tab:Platform Ablation}
\end{table}

\paragraph{Model Component Analysis} 
We analyze the effects of different components in our  model, introducing three variants: one without platform-aware contrastive learning objective , one without node enhancement, and one without platform attention. The experimental results are shown in Table~\ref{tab:Ablation Study}.
From the table, the variants obtain declined performance than the full model, showing the effectiveness of each component. The variant without Node Enhancement struggles to capture platform-specific features, while the one without Platform Attention suffer from significant noise in user comments, leading to suboptimal detection results.

\paragraph{Effect of Modeling Propagation across Different Platforms}
As shown in Table~\ref{tab:Platform Ablation}, we further evaluate the impact of different platform propagation methods.
As removing propagation data on specific platforms, the detection performance obtains different degrees of degradation, particularly when large amounts of propagation data are missing (e.g., on X and Reddit), leading to worse detection than models trained only on news content.

\section{Conclusion}

% \iffalse
% 本文研究了多平台虚假新闻检测的问题。本文认为虚假新闻在不同社交媒体平台上的传播差异对虚假新闻检测有所帮助，为了验证这个结论，本文首先收集了MPPFND Dataset以解决目前单一新闻的不同平台传播的数据集缺失问题。基于构建的数据集的统计特征，本文验证了不同社交媒体平台的传播内容存在差异。此外，本文进一步提出了多平台虚假新闻检测模型MPPFND，以验证社交平台之间的传播内容差异性是否对虚假信息的检测有帮助。通过在数据集上的实验验证了不同社交平台的传播内容差异性对虚假信息检测的效果有提升。
% \fi

This paper investigates fake news detection through propagation across multiple platforms.
To this end, we introduce a new fake news dataset and conduct an empirical study to analyze propagation patterns across different platforms. Through semantic analysis and propagation feature analysis of the dataset we collected, we summarize the commonalities and differences in news propagation across different social media platforms, providing insights and a foundation for designing a fake news detection model across multiple social media platforms. To further demonstrate the contribution of multi-platform propagation data to detection, we compare the detection performance using data propagated on a single social media platform versus data propagated across multiple platforms. The results validate that multi-platform propagation data helps improve the effectiveness of fake news detection.

\section{Acknowledgments}

This work was supported by the National Natural Science Foundation of China (No. U24A20335), the China Postdoctoral Science Foundation (No. 2024M753481), and the Postdoctoral Fellowship Program of China Postdoctoral Science Foundation (No. GZC20232969).

% \section*{Ethics Statement}

% \iffalse
% 道德声明
% \fi

% The dataset does not contain harmful content and does not disclose specific users of comments and posts, thereby protecting users' privacy information. The annotations are collected on a publicly available social media and will be released publicly for future use.

\nocite{ChalnickBillman1988a}
\nocite{Feigenbaum1963a}
\nocite{Hill1983a}
\nocite{OhlssonLangley1985a}
% \nocite{Lewis1978a}
\nocite{Matlock2001}
\nocite{NewellSimon1972a}
\nocite{ShragerLangley1990a}

\bibliographystyle{apacite}

\setlength{\bibleftmargin}{.125in}
\setlength{\bibindent}{-\bibleftmargin}

\bibliography{CogSci_Template}

\end{document}